\documentclass[%
 reprint,
amsmath,amssymb,
aps,
]{revtex4-1}
\usepackage{graphicx}
\usepackage{dcolumn}
\usepackage{bm}
\usepackage{epstopdf}

\begin{document}

\preprint{APS/123-QED}

\title{Disorder-sensitive node-like small gap in FeSe}

\author{Yue Sun,$^1$}
\email{sunyue@issp.u-tokyo.ac.jp}
\author{Shunichiro Kittaka,$^1$ Shota Nakamura,$^1$ Toshiro Sakakibara,$^1$ Peng Zhang,$^1$ Shik Shin,$^1$ Koki Irie,$^2$ Takuya Nomoto,$^3$ Kazushige Machida,$^2$ Jingting Chen,$^4$ and Tsuyoshi Tamegai$^4$}

\affiliation{%
$^1$Institute for Solid State Physics (ISSP), The University of Tokyo, Kashiwa, Chiba 277-8581, Japan\\
$^2$Department of Physics, Ritsumeikan University, Kusatsu, Shiga 525-8577, Japan\\
$^3$RIKEN Center for Emergent Matter Science (CEMS), Hirosawa, Wako, Saitama 351-0198, Japan\\
$^4$Department of Applied Physics, The University of Tokyo, Bunkyo-ku, Tokyo 113-8656, Japan}


\begin{abstract}
We investigate the band structure, nematic state and superconducting gap structure of two selected FeSe single crystals containing different amount of disorder. Transport and angle-resolved photoemission spectroscopy measurements show that the small amount of disorder has little effect to the band structure and the nematic state of FeSe. However, temperature and magnetic field dependencies of specific heat for the two samples are quite different. Wave-vector-dependent gap structure are obtained from the three dimensional field-angle-resolved specific heat measurements. A small gap with two vertical-line nodes or gap minima along the $k_z$ direction is found only in the sample with higher quality. Such symmetry-unprotected nodes or gap minima are found to be smeared out by small amount of disorder, and the gap becomes isotropic in the sample of lower quality. Our study reveals that the reported controversy on the gap structure of FeSe is due to the disorder-sensitive node-like small gap.

\end{abstract}

\maketitle
Among iron-based superconductors (IBSs), FeSe composed of only Fe-Se layers \cite{HsuFongChiFeSediscovery} has the simplest crystal structure, and is usually regarded as the parent compound. It also manifests very intriguing properties including a nematic state without long-range magnetic order \cite{McQueenPRL}, crossover from Bardeen-Cooper-Schrieffer (BCS) to Bose-Einstein-condensation (BEC) \cite{Kasahara18112014}, and a Dirac-cone-like state \cite{ZhangFeSeDirac,KontariDiracconePhysRevLett,SunPhysRevB.93.104502}. Recently, an unexpected high $T_c$ with a sign of superconductivity over 100 K observed in monolayered FeSe \cite{WangCPLMonolayerFeSe,GeNatMatter} makes this system a promising candidate for achieving high-temperature superconductivity and probing the mechanism of superconductivity.

To understand these intriguing properties and the unexpected high $T_c$ in FeSe system, it is crucial to know the gap structure, which is unfortunately still under debate. Both the presence of nodes \cite{SongScience,Kasahara18112014} or nodeless but deep minima \cite{LinFeSeSHPRB,LinJiaoSciRep,hopePhysRevLett,AbdelHcFeSePRB} have been proposed. Even with the same technique, different groups reported different results. In the scanning tunneling microscopy (STM), both the V-shape \cite{SongScience,Kasahara18112014} and U-shape spectrum (finite energy range) \cite{LinJiaoSciRep,BPQIarxiv} were reported. For the thermal conductivity, the difference in the residual $\kappa_0/T$ reported by different groups are more than 20 times \cite{Kasahara18112014,FeSeoldthermalPRB,hopePhysRevLett}. For the electronic specific heat at low temperatures, a linear decrease \cite{LinFeSeSHPRB}, a humplike behavior \cite{LinJiaoSciRep,SunPRBARSHFeSe}, and even a second jump \cite{FeSeSHSecondJumpPhysRevB.96.064524} were observed in previous reports. Such controversy may come from the difference in sample quality since the possible nodes are not symmetry protected as proposed by the theoretical calculation \cite{KreiselPRB}. Experimentally, the change of $T_c$ by disorder has already been confirmed by comparing different crystals \cite{BohmerdisorderPhysRevB.94.024526,RosslerPhysRevB.97.094503} and particle-irradiation effects \cite{FeSeelectronIrra,SunPhysRevBprotonirra,SunPRBheavyionIrradiation}. Therefore, comparison of the gap structure of samples containing different amount of disorder is promising to solve the controversy in gap structure.

A pioneering research using thermal conductivity indeed observed some differences in samples with different qualities \cite{hopePhysRevLett}. Unfortunately, the multigap structure as well as the existence of a very small gap hinder to reveal gap-structure change by traditional temperature or field dependence of quasiparticles (QPs) \cite{hopePhysRevLett}. Similar difficulty also happens in specific heat. Fitting of the temperature dependent electronic specific heat in previous reports concluded different gap structures because too many variables are included by the multigap structure \cite{LinFeSeSHPRB,LinJiaoSciRep,SunPRBARSHFeSe,FeSeSHSecondJumpPhysRevB.96.064524}. To solve this issue and directly observe the changes in gap structure, a technique capable of probing QPs excitations with angular resolution is needed. Field-angle-resolved specific heat (ARSH) measurement is an ideal tool for probing the density of states (DOS) of QPs, and it is angle resolved because the low-lying QP excitations near the gap nodes (minima) are field-orientation dependent \cite{SakakibaraReview}.

In this report, we investigate the band structure, nematic state, and superconducting gap structure of two selected FeSe single crystals containing different amount of disorder. A small amount of disorders was found to have little effect on the band structure and nematic state. On the other hand, the ARSH measurements reveal that the high-quality sample contains a small gap with two vertical-line nodes or gap minima. However, such nodes or gap minima are smeared out by disorder in the lower-quality sample.

\begin{figure*}\center
\includegraphics[width=16cm]{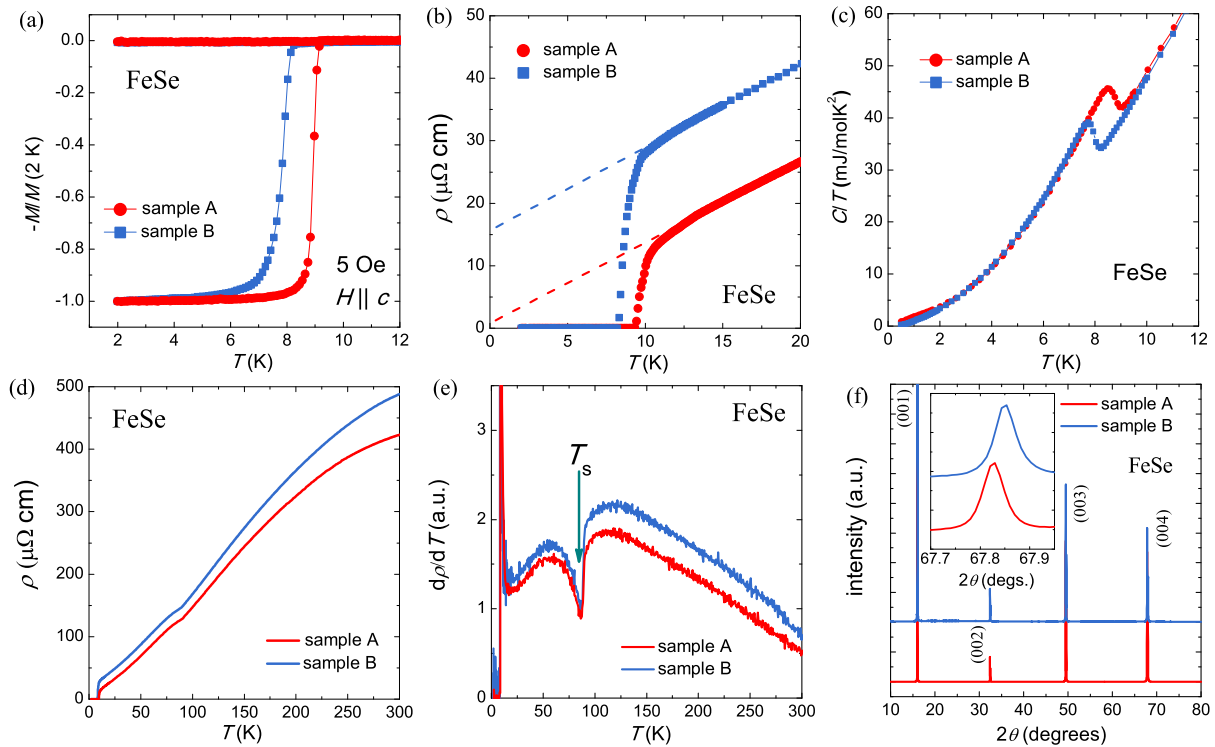}\\
\caption{Temperature dependences of the (a) normalized magnetization at 5 Oe, (b) resistivity at zero field below 20 K, (c) specific heat at zero field, (d) resistivity in the temperature range of 0-300 K, (e) first derivative of $\rho-T$ for the two samples A (red) and B (blue). (f) XRD patterns for sample A and B. The inset is the enlarged part of the (004) peaks.}\label{}
\end{figure*}

FeSe single crystals were grown by the vapor transport method \cite{SunPhysRevBJcFeSe,SunPhysRevB.93.104502}. Two kinds of crystals labeled sample A and sample B were selected from two different batches. Structure of the samples was characterized by means of X-ray diffraction (XRD) with Cu-K$\alpha$ radiation. Chemical compositions were determined by using a scanning electron microscope (SEM) equipped with EDX. Magnetization measurements were performed using a commercial SQUID magnetometer (MPMS-XL5, Quantum Design). Transport measurements were made by using the six-lead method with the applied field parallel to $c$-axis and perpendicular to the applied current in a physical-property-measurement system (PPMS). The ARPES measurements were performed on a spectrometer with a VG-Scienta R4000WAL electron analyzer and a laser delivering 6.994 eV photons. The energy resolution of the system was set to $\sim$5 meV. All measurements were done in ultra high vacuum better that 5$\times$10$^{-11}$ Torr. The temperature dependence of the specific heat was also measured by using PPMS. The magnetic-field dependence of the specific heat was measured in a dilution refrigerator under fields up to 14.7 T. The field-orientation dependence of the specific heat was measured in an 8 T split-pair superconducting magnet with a $^3$He refrigerator. The refrigerator can be continuously rotated by a motor on top of the Dewar with an angular resolution better than 0.01$^\circ$. More details about the calibration and the validity of the measurement system were presented in our review paper \cite{SakakibaraReview}.

Figure 1(a) shows the temperature dependence of the normalized magnetization at 5 Oe for the two samples A and B. Sample A displays $T_c$ $\sim$ 9.2 K, which is slightly higher than that of $\sim$ 8.3 K observed in sample B. The small difference in $T_c$ can be also witnessed in the temperature dependences of resistivity (Fig. 1(b)) and specific heat (Fig. 1(c)). From the low-temperature $\rho-T$ curves in Fig. 1(b), we can obtain the residual resistivity, $\rho_0$, which is directly related to the amount of disorders. $\rho_0$ for the two samples were estimated by linearly extrapolating the normal state data above $T_c$ to $T$ = 0 K as shown by the dashed lines in Fig. 1(b). The $\rho_0$ for sample A is only $\sim$ 1 $\mu\Omega$ cm, which is the smallest among all the reported values \cite{Kasahara18112014,LinJiaoSciRep,hopePhysRevLett}. Such extremely small $\rho_0$ confirms the small amount of disorder as well as the high quality of our sample A. On the other hand, the $\rho_0$ for sample B is $\sim$ 16 $\mu\Omega$ cm. (The simple linear extrapolation may cause some error up to 1 $\mu\Omega$ cm. However, it will not affect the magnitude relation in $\rho_0$ of the two samples.) Together with the resistivity values at 300 K (Fig. 1(d)), the residual resistivity ratio RRR, defined as $\rho$(300 K)/$\rho_0$, is estimated as $\sim$425 for sample A, and $\sim$30 for sample B, respectively. The increase in $\rho_0$ and the decrease in RRR  indicate more disorder in sample B, which is consistent with the slightly lower $T_c$. Although the sample B contains more disorder, both samples manifest relatively sharp superconducting transition width in the magnetization, resistivity, and specific heat (Figs. 1(a) - (c)), which indicates the homogeneous distribution of disorder in sample B. In addition to the difference in $T_c$, the specific heat at $T <$ 2 K are also different for the two samples, which will be discussed in detail later.

Figure 1(d) compares the temperature dependence of resistivity for the two samples up to 300 K. An obvious kink-like behavior below 100 K is observed in both samples, which is related to the structural transition from tetragonal to orthorhombic structure \cite{McQueenPRL}. It can be seen more clearly in the first derivative of the temperature-dependent resistivity d$\rho$/d$T$ as shown in Fig. 1(e). Obviously, the structural transition temperatures, $T_s$, are very close for the two samples (86.4 $\pm$ 0.5 K for sample A, and 86.1 $\pm$ 0.5 K for sample B). Since the structural transition is believed to be driven by the nematic order \cite{NakayamaPRL,WatsonPRB91,WatsonPRB2016,Coldeaannurev,FedorovARPESSciRep}, the similar $T_s$ in samples A and B suggests that such small amount of disorder affects little the nematic order of FeSe. More evidence for the robust nematic order provided by the ARPES measurements will be discussed below.

To get more information about the disorder in both samples, we performed structural and compositional analyses. Fig. 1(f) shows the single crystal XRD pattern for the two samples. Only the (00$l$) peaks can be identified for both samples. Compared to sample A, the positions of peaks are found to be slightly shifted to higher angle in sample B, which can be seen more clearly in the enlarged (004) peaks in the inset. To obtain the lattice constant $a$/$b$, we measured (103) and (104) peaks by scanning the crystal angle $\omega$ independently of 2$\theta$ (angle between incident and scattered X-rays). The lattice constants are estimated as $c$ = (5.524$\pm$0.002) {\AA}, $a$ = (3.777$\pm$0.002) {\AA} for sample A, and $c$ = (5.521$\pm$0.002) {\AA}, $a$ = (3.765$\pm$0.002) {\AA} for sample B. Both $a$ and $c$ decrease in sample B, indicating shrinkage of the lattice. The EDX result shows that the molar ratio of Fe : Se in sample A is $\sim$ 1 : 1.005, which is roughly the stoichiometric FeSe. In the sample B, the ratio is $\sim$ 1 : 1.074, which means the amount of Fe is less than Se. Such changes in the structure and composition with increasing the amount of disorder have also been confirmed in three more kinds of FeSe single crystals with different quality. The shrinkage of lattice and the less amount of Fe suggest that the disorders in sample B could be Fe vacancies, which has been reported by recent STM measurements \cite{LinJiaoSTMFevacancyFeSePhysRevB.96.094504}.

\begin{figure}\center
\includegraphics[width=8.5cm]{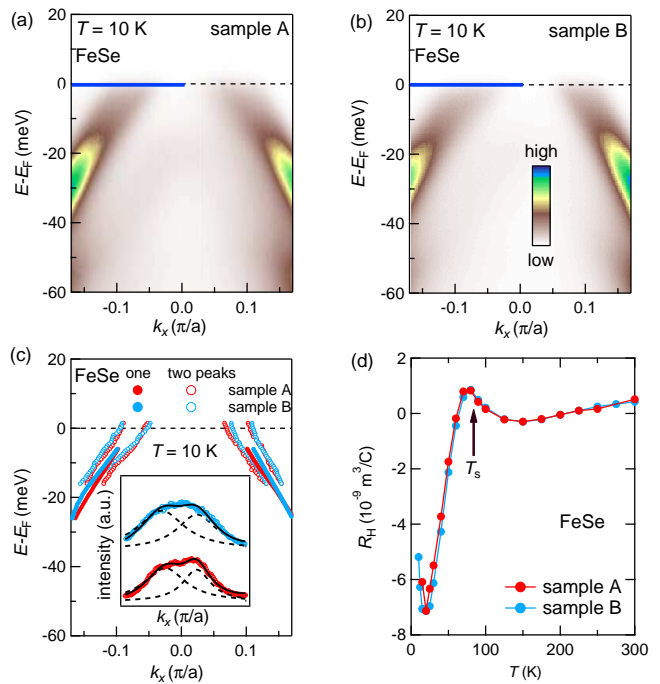}\\
\caption{ARPES intensity plot of (a) sample A and (b) sample B at 10 K measured with $p$-polarized photons. The bands with $d_{yz}$ orbital character are revealed because of the matrix element effect \cite{Hashimotonatcomm}. (c) Extracted band structures from (a) and (b). The band structures are extracted by fitting the momentum distribution curvs (MDCs) with Lorentzian peaks. Fitting function with four Lorentzian peaks are used for the MDCs in the range of -18 meV $\sim$ 0 meV, while fitting function with two Lorentzian peaks are used for the the MDCs in the range of -25 meV $\sim$ -10 meV. The inset shows the details of the fitting on the MDCs indicated by the blue solid lines in (a) and (b). (d) Temperature dependence of Hall coefficients for samples A and B.}\label{}
\end{figure}

Figures 2(a) and (b) show the ARPES intensity plots around $\Gamma$-point for sample A and B at 10 K, respectively. The band splitting due to the nematic electronic order is observed in both samples similar to previous reports \cite{NakayamaPRL,WatsonPRB91,WatsonPRB2016,Coldeaannurev,FedorovARPESSciRep}. As a result of the energy splitting, the band structure was extracted from the momentum distribution curves (MDCs) with the function of four-Lorentzian peaks (MDCs in the range of -18 meV $\leq$ $E$-$E_F$ $\leq$ 0 meV) and that of two-Lorentzian peaks (MDCs in the range of -25 meV $\leq$ $E$-$E_F$ $\leq$ -10 meV). The details of the Lorentzian fitting of the MDCs indicated by the blue solid lines are displayed in Fig. 2(c), where the linewidth of sample B is slightly broader than that of sample A, and its band splitting also becomes obscured. The broader linewidth and the obscured band splitting all suggest that the sample B contains more disorder than sample A, which is consistent with the composition and resistivity results. On the other hand, the change in linewidth is very small, indicating only slightly more disorder in sample B. The obtained band structures of the two samples are compared in Fig. 2(c), which are very similar despite the different degrees of disorder. The ARPES results confirm that the nematic order is little affected, consistent with the unchanged $T_s$.

Figure 2(d) shows the temperature dependences of Hall coefficients ($R_H$) for the two samples, which is determined by the linear part of $\rho_{yx}$ at small field as described in our previous publication \cite{SunPhysRevB.93.104502}. $R_H$ of the two samples are almost identical at temperatures above $T_s$, while they are slightly different at low temperatures. At temperatures above $T_s$, the value of $R_H$ are very small, and close to zero, which can be easily understood by considering a simply compensated two-band model containing equal numbers of electron- and hole-typed charge carriers with similar mobility \cite{SunPhysRevB.93.104502}. Thus, the almost identical $R_H$ for the two samples indicates that the electron- and hole-typed bands are compensated in both samples, which is consistent with the similar band structures observed by ARPES. At temperatures below $T_s$, $R_H$ decreases quickly with temperature, and shows a large negative values, which indicates that the electron-typed charge carriers become dominant. It may come from the increase of mobility of the electron-typed band \cite{SunPhysRevB.93.104502}. The small difference in $R_H$ at low temperatures represents the different amount of disorders of the two samples.

\begin{figure}\center
\includegraphics[width=8cm]{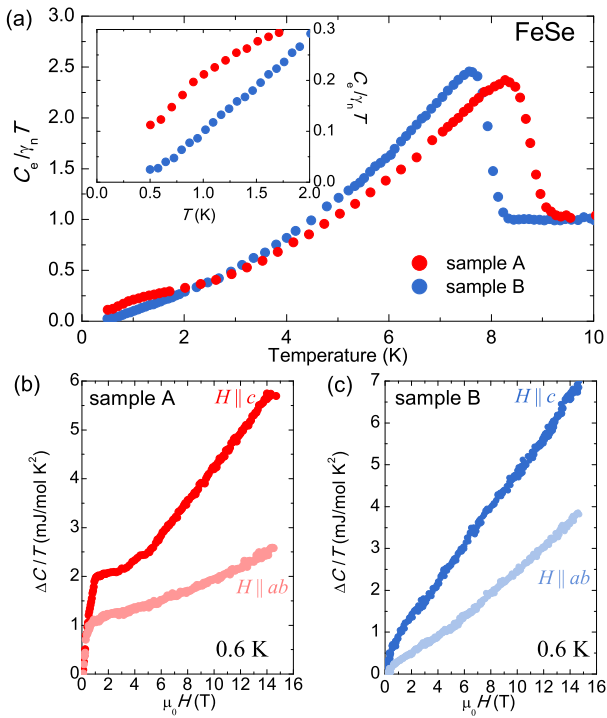}\\
\caption{(a) Normalized zero-field electronic-specific heat, $C_e/\gamma_nT$ vs $T$, for sample A (red) and B (blue). Inset is the enlarged low temperature part. Magnetic-field-induced changes in the specific heat divided by temperature $\Delta C$/$T$ ($\Delta C$ = $C$($H$) - $C$(0T)) for $H\parallel c$ and $H\parallel ab$ at 0.6 K for sample (b) A and (c) B.}\label{}
\end{figure}

To get insight into the effect of disorder on the gap structure, we compare the temperature dependence of zero-field electronic specific heat $C_{\rm e}/T$ of the two samples. As shown in Fig. 3(a), a hump-like behavior below 2 K is observed in sample A, which is typical of multi-gap superconductors like MgB$_2$ and Lu$_2$Fe$_3$Si$_5$ \cite{MgB2PhysRevLetttwogap,*NakajimaRuFeSitwogapPhysRevLett.100.157001}. However, such hump-like behavior is strongly suppressed in sample B, which can be seen more clearly in the enlarged low $T$ part in the inset. A similar result has recently been reported in Ref. \cite{RosslerPhysRevB.97.094503}. Both hump-like and linear behavior have been observed in previous reports \cite{LinFeSeSHPRB,LinJiaoSciRep,SunPRBARSHFeSe,RosslerPhysRevB.97.094503,Hardyarxiv}. Our results here directly prove that such controversy originates from the sample-dependent disorder. Besides, the large divergence in $C_{\rm e}$ at low $T$ between the two samples indicates the differences in their gap structure.

\begin{figure*}\center
\includegraphics[width=16cm]{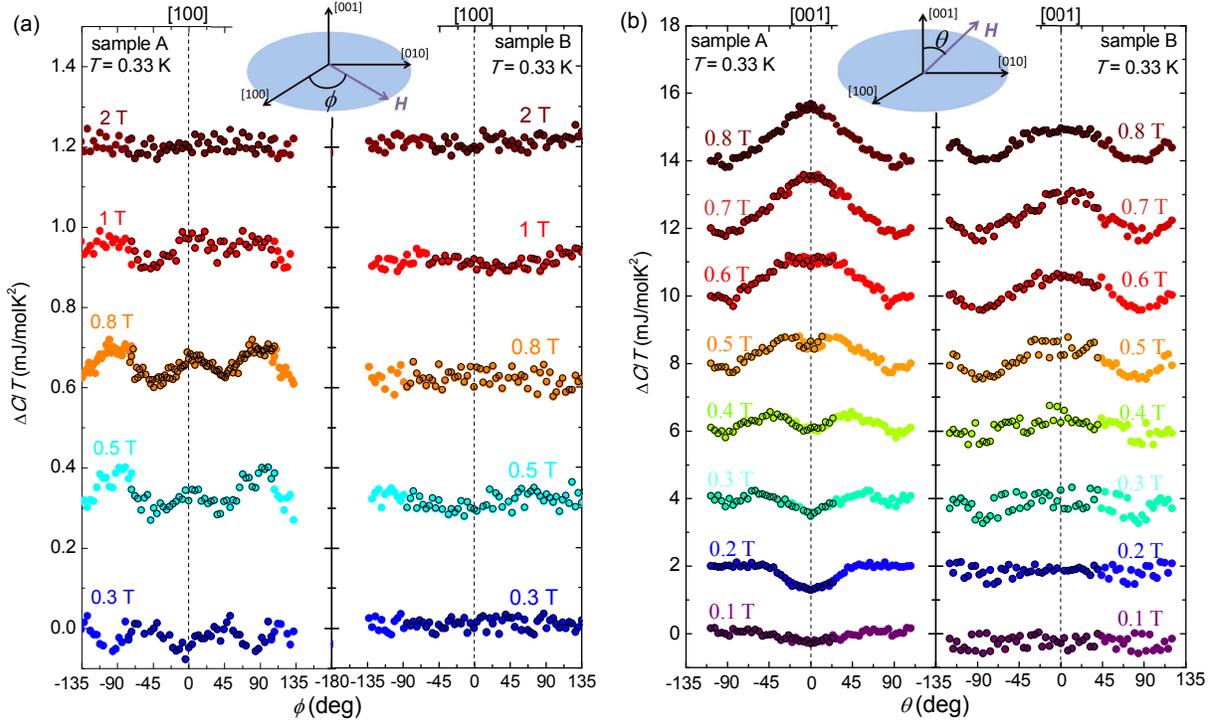}\\
\caption{(a) Azimuthal angle dependence of the specific heat $\Delta C(\phi)/T$ measured under various fields at 0.33 K for samples A (left panel) and B (right panel). $\Delta C(\phi)/T$ is defined as $C(\phi)/T$-$C(-45^\circ)/T$, and each subsequent curve is shifted vertically by 0.3 mJ/molK$^2$. (b) Polar angle dependence of the specific heat $\Delta C(\theta)/T$ measured under various fields at 0.33 K for samples A (left panel) and B (right panel). $\Delta C(\theta)/T$ is defined as $C(\theta)/T$-$C(-90^\circ)/T$, and each subsequent curve is shifted vertically by 2 mJ/molK$^2$. Black-outlined symbols are the measured data; the others are mirrored points to show the symmetry more clearly. The in-plane (azimuthal) angle $\phi$ is defined as the angle away from [100] direction, while the out-of-plane (polar) angle $\theta$ is defined as the angle away from [001] direction as shown in the insets of (a) and (b), respectively.}\label{}
\end{figure*}

More information about the gap structure can be obtained from the magnetic field dependence of the specific heat, $C/T$ vs $H$, which reflects QP excitations across the SC gap.
Changes in the specific heat, $\Delta C$/$T$ ($\Delta C$ = $C$($H$) - $C$(0 T)), of the two samples at 0.6 K for $H \parallel c$ and $H \parallel  ab$ are shown in Figs. 3(b) and (c), respectively. For sample A, $\Delta C$/$T$ first increases rapidly with the same slope for both $H\parallel c$ and $H\parallel ab$ up to $\sim$1 mJ/molK$^2$, which indicates the presence of a small gap with little out-of-plane anisotropy in the small gap. As $H$ further increases, $\Delta C$/$T$ changes with different slopes for $H\parallel c$ and $H\parallel ab$. The $\Delta C(H\parallel c)/T$ even changes slopes twice indicating the multigap-structure, which has been discussed in detail in our previous report \cite{SunPRBARSHFeSe}. On the other hand, the slope of $\Delta C$/$T$ is almost constant at higher fields, but with different values depending on the field direction in sample B, which indicates that the larger gap has an appreciable anisotropy. The initial steeper increase of $\Delta C$/$T$ at low fields also depends on the  field direction, which indicates that the smaller gap is also anisotropic. Similar behavior of $C/T$ vs $H$ to that in sample B has also been reported in previous publications \cite{LinFeSeSHPRB,SatoFeSeSPNAS}. The large differences in temperature and magnetic field dependences of specific heat of the two samples suggest that the superconducting gap structure is very sensitive to the disorder although the band structure and nematic state are little affected.

To resolve the differences in gap structures of the two samples, we turn to the measurements of the specific heat under magnetic field with angle resolution. Due to the  Doppler shift, $\delta E = m_e \textbf{\emph{v}}_F \cdot \textbf{\emph{v}}_s$ ($m_e$ is the electron mass,  $\emph{\textbf{v}}_F$ is the Fermi velocity, and $\emph{\textbf{v}}_s$ is the local superfluid velocity perpendicular to the field) \cite{VolovikJETPLett}, the zero-energy DOS in the vortex core in superconductors with nodes (or minima) depends on the direction of the field with respect to the nodal position. When $H\parallel$ node, it shows minima because $\delta E=0$ in the case of $\emph{\textbf{v}}_F\perp\emph{\textbf{v}}_s$, while, it turns to maxima when $H\perp$ node because $\delta E$ becomes maximal in the situation of $\emph{\textbf{v}}_F\parallel\emph{\textbf{v}}_s$. Therefore, the specific heat shows symmetric oscillation under the rotating magnetic field. By contrast, for superconductors with isotropic gap, the specific heat should be independent of the field direction.

Figure 4(a) compares the azimuthal angle-resolved $\Delta C(\phi)/T$ at 0.33 K for samples A (left panel) and B (right panel). For sample A, $\Delta C(\phi)/T$ manifests an obvious four-fold symmetry. Below 0.5 T, $\Delta C(\phi)/T$ shows minima for $H\|[100]$ and $H\|[010]$ ($\phi=0^\circ$ and $90^\circ$) and maxima for the $H\|[110]$ ($\phi=45^\circ$). At $H \geq $ 0.5 T, $\Delta C(\phi)/T$ becomes maxima for the $H\|[100]$ and $H\|[010]$, but minima for the $H\|[110]$. Such sign-change behavior in the oscillation is due to the strong enhancement of QP scattering under relatively large magnetic field. In this case, a much higher finite-energy DOS around the nodal position will be excited for $H\parallel$ nodes. When the finite-energy DOS overcomes the zero-energy DOS, the oscillation switches signs \cite{VorontsovPhysRevLett.96.237001,HiragiJPSJASHcal}. Such a sign change is commonly observed in superconductors with nodes \cite{AnPhysRevLett.104.037002,KittakaKFe2As2JPSJ} or gap minima \cite{KittakaCeRu2JPSJ.82.123706}. Thus, the four-fold symmetric $\Delta C(\phi)/T$ observed in sample A demonstrates the existence of nodes or gap minima. On the other hand, the $\Delta C(\phi)/T$ for sample B shows no oscillation, which indicates that the gap is almost isotropic in the measured temperature and field range.

The difference in gap structure can be witnessed in the polar-angle dependence of the specific heat $\Delta C(\theta)/T$ as shown in Fig. 4(b). For sample A, $\Delta C(\theta)/T$ first shows minima in the [001] direction ($H\parallel c$) with two shoulders under small fields. With increasing field, the minima gradually increase, and the two shoulders move towards the [001] direction. Finally, the minima at [001] turn to maxima at $\sim$0.8 T. The anisotropy-inverted $\Delta C(\theta)/T$ can also be explained by the competition between the zero-energy and finite-energy DOSs based on the Doppler shift. In the case of vertical-line nodes (or gap minima), $\textbf{\emph{v}}_F^{H\parallel c}\cdot\textbf{\emph{v}}_s^{H\parallel c}<\textbf{\emph{v}}_F^{H\parallel ab}\cdot\textbf{\emph{v}}_s^{H\parallel ab}$ in the small-field region because of Fermi-surface warping along $k_z$-direction. At higher fields, the scattering of QPs is largely enhanced for $H\parallel$ nodal (or gap minima) lines, making $\textbf{\emph{v}}_F^{H\parallel c}\cdot\textbf{\emph{v}}_s^{H\parallel c}>\textbf{\emph{v}}_F^{H\parallel ab}\cdot\textbf{\emph{v}}_s^{H\parallel ab}$. Considering the existence of twin boundaries in the orthorhombic phase of FeSe, the oscillations observed in the azimuthal and polar angle-resolved specific heat results in sample A have been proved to originate from a small gap ($\sim$ 0.39 meV) with two vertical-line nodes or gap minima along the $k_z$ direction \cite{SunPRBARSHFeSe} (Schematic gap structure is shown in Fig. 3(d) of Ref. \cite{SunPRBARSHFeSe}). The existence of a very small gap has also been reported in other specific heat \cite{LinFeSeSHPRB,LinJiaoSciRep} or thermal conductivity measurements of QPs \cite{hopePhysRevLett}. On the other hand, $\Delta C(\theta)/T$ of sample B only shows a two-fold symmetry without shape change, which simply reflects the out-of-plane anisotropy of $H_{c2}$. It again confirms the almost in-plane isotropic gap structure at low temperatures in sample B.

The above comparison of the 3D ARSH for the two samples clearly proves that the nodes or gap minima in the small gap of sample A has been smeared and turns to be almost isotropic in sample B. Although the amount of disorders is too small to affect the band structure and nematic order, it dramatically changes the gap structure. It is due to the symmetry-unprotected nature of the nodes (or gap minima) as well as the small size of the gap ($\sim$ 0.39 meV), which are sensitive to disorder. Therefore, the reported controversy on the gap structure of FeSe by different groups on different samples now can be understood by the presence of disorder-sensitive node-like small gap. In the clean sample similar to the sample A, the node-like small gap will contribute more QPs at low temperatures under small field, which causes the larger value of $C_e$ (inset of Fig. 3(a)) and the faster increase in $C/T$ vs $H$ (Fig. 3(b)). On the other hand, in a slightly dirtier sample like the sample B, the nodes or gap minima in the small gap will be suppressed by averaging of gap values due to disorder. It will make the excitations of QPs at low temperatures and/or low fields more difficult, leading to the suppression of the value of $C_e$ and reduction of the field-induced change in  $C_e/T$. Together with the smearing of the nodes or gap minima, size of the gap should be also reduced, which is supported by the slightly lower $T_c$ in sample B. The above discussion is also consistent with the report on the H$^+$-irradiation effect in FeSe \cite{SunPhysRevBprotonirra}. Since the size of the node-like gap is very small, we cannot exclude the extreme possibility that it has been totally suppressed in sample B. Considering the small gap may come from a tiny band, it could not be observed in ARPES measurement because of the limit of resolution. On the other hand, the anisotropic larger gap in FeSe as reported by the STM \cite{BPQIarxiv} is hard to be smeared by such a small amount of disorder. Thus, the whole gap structure of sample B is still anisotropic, and only the small gap becomes isotropic.

In summary, we systematically studied the structure, composition, superconductivity, transport, band structure, nematic order, and the gap structure of two selected FeSe single crystals containing different amounts of disorder. Small amount of disorder has been proved to affect little the band structure, and the nematic order. However, the temperature and magnetic field dependences of specific heat of the two samples have been found to be quite different. 3D ARSH measurements demonstrate the presence of a small gap with two vertical-line nodes or gap minima along the $k_z$ direction in the sample with higher quality. Such symmetry-unprotected nodes or gap minima are smeared by disorder and the gap becomes almost isotropic in the sample with lower quality. Our study clearly reveals that the reported controversy on the gap structure of FeSe is due to the presence of the disorder-sensitive node-like small gap.

T. N. is supported by RIKEN Special Postdoctoral Researchers Program. The present work was supported by a Grant-in-Aid for Scientific Research on Innovative Areas ``J-Physics'' (15H05883) from MEXT, and KAKENHI (17K05553, 15K05158, and 17H01141) from JSPS.


\bibliography{references}

\begin{thebibliography}{42}%
\makeatletter
\providecommand \@ifxundefined [1]{%
 \@ifx{#1\undefined}
}%
\providecommand \@ifnum [1]{%
 \ifnum #1\expandafter \@firstoftwo
 \else \expandafter \@secondoftwo
 \fi
}%
\providecommand \@ifx [1]{%
 \ifx #1\expandafter \@firstoftwo
 \else \expandafter \@secondoftwo
 \fi
}%
\providecommand \natexlab [1]{#1}%
\providecommand \enquote  [1]{``#1''}%
\providecommand \bibnamefont  [1]{#1}%
\providecommand \bibfnamefont [1]{#1}%
\providecommand \citenamefont [1]{#1}%
\providecommand \href@noop [0]{\@secondoftwo}%
\providecommand \href [0]{\begingroup \@sanitize@url \@href}%
\providecommand \@href[1]{\@@startlink{#1}\@@href}%
\providecommand \@@href[1]{\endgroup#1\@@endlink}%
\providecommand \@sanitize@url [0]{\catcode `\\12\catcode `\$12\catcode
  `\&12\catcode `\#12\catcode `\^12\catcode `\_12\catcode `\%12\relax}%
\providecommand \@@startlink[1]{}%
\providecommand \@@endlink[0]{}%
\providecommand \url  [0]{\begingroup\@sanitize@url \@url }%
\providecommand \@url [1]{\endgroup\@href {#1}{\urlprefix }}%
\providecommand \urlprefix  [0]{URL }%
\providecommand \Eprint [0]{\href }%
\providecommand \doibase [0]{http://dx.doi.org/}%
\providecommand \selectlanguage [0]{\@gobble}%
\providecommand \bibinfo  [0]{\@secondoftwo}%
\providecommand \bibfield  [0]{\@secondoftwo}%
\providecommand \translation [1]{[#1]}%
\providecommand \BibitemOpen [0]{}%
\providecommand \bibitemStop [0]{}%
\providecommand \bibitemNoStop [0]{.\EOS\space}%
\providecommand \EOS [0]{\spacefactor3000\relax}%
\providecommand \BibitemShut  [1]{\csname bibitem#1\endcsname}%
\let\auto@bib@innerbib\@empty
\bibitem [{\citenamefont {Hsu}\ \emph {et~al.}(2008)\citenamefont {Hsu},
  \citenamefont {Luo}, \citenamefont {Yeh}, \citenamefont {Chen}, \citenamefont
  {Huang}, \citenamefont {Wu}, \citenamefont {Lee}, \citenamefont {Huang},
  \citenamefont {Chu}, \citenamefont {Yan},\ and\ \citenamefont
  {Wu}}]{HsuFongChiFeSediscovery}%
  \BibitemOpen
  \bibfield  {author} {\bibinfo {author} {\bibfnamefont {F.~C.}\ \bibnamefont
  {Hsu}}, \bibinfo {author} {\bibfnamefont {J.~Y.}\ \bibnamefont {Luo}},
  \bibinfo {author} {\bibfnamefont {K.~W.}\ \bibnamefont {Yeh}}, \bibinfo
  {author} {\bibfnamefont {T.~K.}\ \bibnamefont {Chen}}, \bibinfo {author}
  {\bibfnamefont {T.~W.}\ \bibnamefont {Huang}}, \bibinfo {author}
  {\bibfnamefont {P.~M.}\ \bibnamefont {Wu}}, \bibinfo {author} {\bibfnamefont
  {Y.~C.}\ \bibnamefont {Lee}}, \bibinfo {author} {\bibfnamefont {Y.-L.}\
  \bibnamefont {Huang}}, \bibinfo {author} {\bibfnamefont {Y.-Y.}\ \bibnamefont
  {Chu}}, \bibinfo {author} {\bibfnamefont {D.~C.}\ \bibnamefont {Yan}}, \ and\
  \bibinfo {author} {\bibfnamefont {M.~K.}\ \bibnamefont {Wu}},\ }\href@noop {}
  {\bibfield  {journal} {\bibinfo  {journal} {Proc. Nat. Acad. Sci.}\ }\textbf
  {\bibinfo {volume} {105}},\ \bibinfo {pages} {14262} (\bibinfo {year}
  {2008})}\BibitemShut {NoStop}%
\bibitem [{\citenamefont {McQueen}\ \emph {et~al.}(2009)\citenamefont
  {McQueen}, \citenamefont {Williams}, \citenamefont {Stephens}, \citenamefont
  {Tao}, \citenamefont {Zhu}, \citenamefont {Ksenofontov}, \citenamefont
  {Casper}, \citenamefont {Felser},\ and\ \citenamefont {Cava}}]{McQueenPRL}%
  \BibitemOpen
  \bibfield  {author} {\bibinfo {author} {\bibfnamefont {T.~M.}\ \bibnamefont
  {McQueen}}, \bibinfo {author} {\bibfnamefont {A.~J.}\ \bibnamefont
  {Williams}}, \bibinfo {author} {\bibfnamefont {P.~W.}\ \bibnamefont
  {Stephens}}, \bibinfo {author} {\bibfnamefont {J.}~\bibnamefont {Tao}},
  \bibinfo {author} {\bibfnamefont {Y.}~\bibnamefont {Zhu}}, \bibinfo {author}
  {\bibfnamefont {V.}~\bibnamefont {Ksenofontov}}, \bibinfo {author}
  {\bibfnamefont {F.}~\bibnamefont {Casper}}, \bibinfo {author} {\bibfnamefont
  {C.}~\bibnamefont {Felser}}, \ and\ \bibinfo {author} {\bibfnamefont {R.~J.}\
  \bibnamefont {Cava}},\ }\href@noop {} {\bibfield  {journal} {\bibinfo
  {journal} {Phys. Rev. Lett.}\ }\textbf {\bibinfo {volume} {103}},\ \bibinfo
  {pages} {057002} (\bibinfo {year} {2009})}\BibitemShut {NoStop}%
\bibitem [{\citenamefont {Kasahara}\ \emph {et~al.}(2014)\citenamefont
  {Kasahara}, \citenamefont {Watashige}, \citenamefont {Hanaguri},
  \citenamefont {Kohsaka}, \citenamefont {Yamashita}, \citenamefont
  {Shimoyama}, \citenamefont {Mizukami}, \citenamefont {Endo}, \citenamefont
  {Ikeda}, \citenamefont {Aoyama}, \citenamefont {Terashima}, \citenamefont
  {Uji}, \citenamefont {Wolf}, \citenamefont {von L\"{o}hneysen}, \citenamefont
  {Shibauchi},\ and\ \citenamefont {Matsuda}}]{Kasahara18112014}%
  \BibitemOpen
  \bibfield  {author} {\bibinfo {author} {\bibfnamefont {S.}~\bibnamefont
  {Kasahara}}, \bibinfo {author} {\bibfnamefont {T.}~\bibnamefont {Watashige}},
  \bibinfo {author} {\bibfnamefont {T.}~\bibnamefont {Hanaguri}}, \bibinfo
  {author} {\bibfnamefont {Y.}~\bibnamefont {Kohsaka}}, \bibinfo {author}
  {\bibfnamefont {T.}~\bibnamefont {Yamashita}}, \bibinfo {author}
  {\bibfnamefont {Y.}~\bibnamefont {Shimoyama}}, \bibinfo {author}
  {\bibfnamefont {Y.}~\bibnamefont {Mizukami}}, \bibinfo {author}
  {\bibfnamefont {R.}~\bibnamefont {Endo}}, \bibinfo {author} {\bibfnamefont
  {H.}~\bibnamefont {Ikeda}}, \bibinfo {author} {\bibfnamefont
  {K.}~\bibnamefont {Aoyama}}, \bibinfo {author} {\bibfnamefont
  {T.}~\bibnamefont {Terashima}}, \bibinfo {author} {\bibfnamefont
  {S.}~\bibnamefont {Uji}}, \bibinfo {author} {\bibfnamefont {T.}~\bibnamefont
  {Wolf}}, \bibinfo {author} {\bibfnamefont {H.}~\bibnamefont {von
  L\"{o}hneysen}}, \bibinfo {author} {\bibfnamefont {T.}~\bibnamefont
  {Shibauchi}}, \ and\ \bibinfo {author} {\bibfnamefont {Y.}~\bibnamefont
  {Matsuda}},\ }\href {\doibase 10.1073/pnas.1413477111} {\bibfield  {journal}
  {\bibinfo  {journal} {Proc. Nat. Acad. Sci.}\ }\textbf {\bibinfo {volume}
  {111}},\ \bibinfo {pages} {16309} (\bibinfo {year} {2014})}\BibitemShut
  {NoStop}%
\bibitem [{\citenamefont {Zhang}\ \emph {et~al.}(2016)\citenamefont {Zhang},
  \citenamefont {Yi}, \citenamefont {Liu}, \citenamefont {Li}, \citenamefont
  {Lee}, \citenamefont {Moore}, \citenamefont {Hashimoto}, \citenamefont
  {Nakajima}, \citenamefont {Eisaki}, \citenamefont {Mo}, \citenamefont
  {Hussain}, \citenamefont {Devereaux}, \citenamefont {Shen},\ and\
  \citenamefont {Lu}}]{ZhangFeSeDirac}%
  \BibitemOpen
  \bibfield  {author} {\bibinfo {author} {\bibfnamefont {Y.}~\bibnamefont
  {Zhang}}, \bibinfo {author} {\bibfnamefont {M.}~\bibnamefont {Yi}}, \bibinfo
  {author} {\bibfnamefont {Z.~K.}\ \bibnamefont {Liu}}, \bibinfo {author}
  {\bibfnamefont {W.}~\bibnamefont {Li}}, \bibinfo {author} {\bibfnamefont
  {J.~J.}\ \bibnamefont {Lee}}, \bibinfo {author} {\bibfnamefont {R.~G.}\
  \bibnamefont {Moore}}, \bibinfo {author} {\bibfnamefont {M.}~\bibnamefont
  {Hashimoto}}, \bibinfo {author} {\bibfnamefont {M.}~\bibnamefont {Nakajima}},
  \bibinfo {author} {\bibfnamefont {H.}~\bibnamefont {Eisaki}}, \bibinfo
  {author} {\bibfnamefont {S.~K.}\ \bibnamefont {Mo}}, \bibinfo {author}
  {\bibfnamefont {Z.}~\bibnamefont {Hussain}}, \bibinfo {author} {\bibfnamefont
  {T.~P.}\ \bibnamefont {Devereaux}}, \bibinfo {author} {\bibfnamefont {Z.~X.}\
  \bibnamefont {Shen}}, \ and\ \bibinfo {author} {\bibfnamefont {D.~H.}\
  \bibnamefont {Lu}},\ }\href@noop {} {\bibfield  {journal} {\bibinfo
  {journal} {Phys. Rev. B}\ }\textbf {\bibinfo {volume} {94}},\ \bibinfo
  {pages} {115153} (\bibinfo {year} {2016})}\BibitemShut {NoStop}%
\bibitem [{\citenamefont {Onari}\ \emph {et~al.}(2016)\citenamefont {Onari},
  \citenamefont {Yamakawa},\ and\ \citenamefont
  {Kontani}}]{KontariDiracconePhysRevLett}%
  \BibitemOpen
  \bibfield  {author} {\bibinfo {author} {\bibfnamefont {S.}~\bibnamefont
  {Onari}}, \bibinfo {author} {\bibfnamefont {Y.}~\bibnamefont {Yamakawa}}, \
  and\ \bibinfo {author} {\bibfnamefont {H.}~\bibnamefont {Kontani}},\
  }\href@noop {} {\bibfield  {journal} {\bibinfo  {journal} {Phys. Rev. Lett.}\
  }\textbf {\bibinfo {volume} {116}},\ \bibinfo {pages} {227001} (\bibinfo
  {year} {2016})}\BibitemShut {NoStop}%
\bibitem [{\citenamefont {Sun}\ \emph {et~al.}(2016)\citenamefont {Sun},
  \citenamefont {Pyon},\ and\ \citenamefont {Tamegai}}]{SunPhysRevB.93.104502}%
  \BibitemOpen
  \bibfield  {author} {\bibinfo {author} {\bibfnamefont {Y.}~\bibnamefont
  {Sun}}, \bibinfo {author} {\bibfnamefont {S.}~\bibnamefont {Pyon}}, \ and\
  \bibinfo {author} {\bibfnamefont {T.}~\bibnamefont {Tamegai}},\ }\href
  {\doibase 10.1103/PhysRevB.93.104502} {\bibfield  {journal} {\bibinfo
  {journal} {Phys. Rev. B}\ }\textbf {\bibinfo {volume} {93}},\ \bibinfo
  {pages} {104502} (\bibinfo {year} {2016})}\BibitemShut {NoStop}%
\bibitem [{\citenamefont {Wang}\ \emph {et~al.}(2012)\citenamefont {Wang},
  \citenamefont {Li}, \citenamefont {Zhang}, \citenamefont {Zhang},
  \citenamefont {Zhang}, \citenamefont {Li}, \citenamefont {Ding},
  \citenamefont {Ou}, \citenamefont {Deng}, \citenamefont {Chang},
  \citenamefont {Wen}, \citenamefont {Song}, \citenamefont {He}, \citenamefont
  {Jia}, \citenamefont {Ji}, \citenamefont {Wang}, \citenamefont {Wang},
  \citenamefont {Chen}, \citenamefont {Ma},\ and\ \citenamefont
  {Xue}}]{WangCPLMonolayerFeSe}%
  \BibitemOpen
  \bibfield  {author} {\bibinfo {author} {\bibfnamefont {Q.~Y.}\ \bibnamefont
  {Wang}}, \bibinfo {author} {\bibfnamefont {Z.}~\bibnamefont {Li}}, \bibinfo
  {author} {\bibfnamefont {W.~H.}\ \bibnamefont {Zhang}}, \bibinfo {author}
  {\bibfnamefont {Z.~C.}\ \bibnamefont {Zhang}}, \bibinfo {author}
  {\bibfnamefont {J.~S.}\ \bibnamefont {Zhang}}, \bibinfo {author}
  {\bibfnamefont {W.}~\bibnamefont {Li}}, \bibinfo {author} {\bibfnamefont
  {H.}~\bibnamefont {Ding}}, \bibinfo {author} {\bibfnamefont {Y.~B.}\
  \bibnamefont {Ou}}, \bibinfo {author} {\bibfnamefont {P.}~\bibnamefont
  {Deng}}, \bibinfo {author} {\bibfnamefont {K.}~\bibnamefont {Chang}},
  \bibinfo {author} {\bibfnamefont {J.}~\bibnamefont {Wen}}, \bibinfo {author}
  {\bibfnamefont {C.~L.}\ \bibnamefont {Song}}, \bibinfo {author}
  {\bibfnamefont {K.}~\bibnamefont {He}}, \bibinfo {author} {\bibfnamefont
  {J.~F.}\ \bibnamefont {Jia}}, \bibinfo {author} {\bibfnamefont {S.~H.}\
  \bibnamefont {Ji}}, \bibinfo {author} {\bibfnamefont {Y.~Y.}\ \bibnamefont
  {Wang}}, \bibinfo {author} {\bibfnamefont {L.~L.}\ \bibnamefont {Wang}},
  \bibinfo {author} {\bibfnamefont {X.}~\bibnamefont {Chen}}, \bibinfo {author}
  {\bibfnamefont {X.~C.}\ \bibnamefont {Ma}}, \ and\ \bibinfo {author}
  {\bibfnamefont {Q.~K.}\ \bibnamefont {Xue}},\ }\href@noop {} {\bibfield
  {journal} {\bibinfo  {journal} {Chin. Phys. Lett.}\ }\textbf {\bibinfo
  {volume} {29}},\ \bibinfo {pages} {037402} (\bibinfo {year}
  {2012})}\BibitemShut {NoStop}%
\bibitem [{\citenamefont {Ge}\ \emph {et~al.}(2015)\citenamefont {Ge},
  \citenamefont {Liu}, \citenamefont {Liu}, \citenamefont {Gao}, \citenamefont
  {Qian}, \citenamefont {Xue}, \citenamefont {Liu},\ and\ \citenamefont
  {Jia}}]{GeNatMatter}%
  \BibitemOpen
  \bibfield  {author} {\bibinfo {author} {\bibfnamefont {J.~F.}\ \bibnamefont
  {Ge}}, \bibinfo {author} {\bibfnamefont {Z.~L.}\ \bibnamefont {Liu}},
  \bibinfo {author} {\bibfnamefont {C.}~\bibnamefont {Liu}}, \bibinfo {author}
  {\bibfnamefont {C.~L.}\ \bibnamefont {Gao}}, \bibinfo {author} {\bibfnamefont
  {D.}~\bibnamefont {Qian}}, \bibinfo {author} {\bibfnamefont {Q.~K.}\
  \bibnamefont {Xue}}, \bibinfo {author} {\bibfnamefont {Y.}~\bibnamefont
  {Liu}}, \ and\ \bibinfo {author} {\bibfnamefont {J.~F.}\ \bibnamefont
  {Jia}},\ }\href@noop {} {\bibfield  {journal} {\bibinfo  {journal} {Nat.
  Mater.}\ }\textbf {\bibinfo {volume} {14}},\ \bibinfo {pages} {285} (\bibinfo
  {year} {2015})}\BibitemShut {NoStop}%
\bibitem [{\citenamefont {Song}\ \emph {et~al.}(2011)\citenamefont {Song},
  \citenamefont {Wang}, \citenamefont {Cheng}, \citenamefont {Jiang},
  \citenamefont {Li}, \citenamefont {Zhang}, \citenamefont {Li}, \citenamefont
  {He}, \citenamefont {Wang}, \citenamefont {Jia}, \citenamefont {Hung},
  \citenamefont {Wu}, \citenamefont {Ma}, \citenamefont {Chen},\ and\
  \citenamefont {Xue}}]{SongScience}%
  \BibitemOpen
  \bibfield  {author} {\bibinfo {author} {\bibfnamefont {C.-L.}\ \bibnamefont
  {Song}}, \bibinfo {author} {\bibfnamefont {Y.-L.}\ \bibnamefont {Wang}},
  \bibinfo {author} {\bibfnamefont {P.}~\bibnamefont {Cheng}}, \bibinfo
  {author} {\bibfnamefont {Y.-P.}\ \bibnamefont {Jiang}}, \bibinfo {author}
  {\bibfnamefont {W.}~\bibnamefont {Li}}, \bibinfo {author} {\bibfnamefont
  {T.}~\bibnamefont {Zhang}}, \bibinfo {author} {\bibfnamefont
  {Z.}~\bibnamefont {Li}}, \bibinfo {author} {\bibfnamefont {K.}~\bibnamefont
  {He}}, \bibinfo {author} {\bibfnamefont {L.}~\bibnamefont {Wang}}, \bibinfo
  {author} {\bibfnamefont {J.-F.}\ \bibnamefont {Jia}}, \bibinfo {author}
  {\bibfnamefont {H.-H.}\ \bibnamefont {Hung}}, \bibinfo {author}
  {\bibfnamefont {C.}~\bibnamefont {Wu}}, \bibinfo {author} {\bibfnamefont
  {X.}~\bibnamefont {Ma}}, \bibinfo {author} {\bibfnamefont {X.}~\bibnamefont
  {Chen}}, \ and\ \bibinfo {author} {\bibfnamefont {Q.-K.}\ \bibnamefont
  {Xue}},\ }\href@noop {} {\bibfield  {journal} {\bibinfo  {journal} {Science}\
  }\textbf {\bibinfo {volume} {332}},\ \bibinfo {pages} {1410} (\bibinfo {year}
  {2011})}\BibitemShut {NoStop}%
\bibitem [{\citenamefont {Lin}\ \emph {et~al.}(2011)\citenamefont {Lin},
  \citenamefont {Hsieh}, \citenamefont {Chareev}, \citenamefont {Vasiliev},
  \citenamefont {Parsons},\ and\ \citenamefont {Yang}}]{LinFeSeSHPRB}%
  \BibitemOpen
  \bibfield  {author} {\bibinfo {author} {\bibfnamefont {J.~Y.}\ \bibnamefont
  {Lin}}, \bibinfo {author} {\bibfnamefont {Y.~S.}\ \bibnamefont {Hsieh}},
  \bibinfo {author} {\bibfnamefont {D.~A.}\ \bibnamefont {Chareev}}, \bibinfo
  {author} {\bibfnamefont {A.~N.}\ \bibnamefont {Vasiliev}}, \bibinfo {author}
  {\bibfnamefont {Y.}~\bibnamefont {Parsons}}, \ and\ \bibinfo {author}
  {\bibfnamefont {H.~D.}\ \bibnamefont {Yang}},\ }\href@noop {} {\bibfield
  {journal} {\bibinfo  {journal} {Phys. Rev. B}\ }\textbf {\bibinfo {volume}
  {84}},\ \bibinfo {pages} {220507} (\bibinfo {year} {2011})}\BibitemShut
  {NoStop}%
\bibitem [{\citenamefont {Lin}\ \emph {et~al.}(2017)\citenamefont {Lin},
  \citenamefont {Huang}, \citenamefont {R\"{o}{\ss}ler}, \citenamefont {Koz},
  \citenamefont {R\"{o}{\ss}ler}, \citenamefont {Schwarz},\ and\ \citenamefont
  {Wirth}}]{LinJiaoSciRep}%
  \BibitemOpen
  \bibfield  {author} {\bibinfo {author} {\bibfnamefont {J.}~\bibnamefont
  {Lin}}, \bibinfo {author} {\bibfnamefont {C.}~\bibnamefont {Huang}}, \bibinfo
  {author} {\bibfnamefont {S.}~\bibnamefont {R\"{o}{\ss}ler}}, \bibinfo
  {author} {\bibfnamefont {C.}~\bibnamefont {Koz}}, \bibinfo {author}
  {\bibfnamefont {U.~K.}\ \bibnamefont {R\"{o}{\ss}ler}}, \bibinfo {author}
  {\bibfnamefont {U.}~\bibnamefont {Schwarz}}, \ and\ \bibinfo {author}
  {\bibfnamefont {S.}~\bibnamefont {Wirth}},\ }\href@noop {} {\bibfield
  {journal} {\bibinfo  {journal} {Sci. Rep.}\ }\textbf {\bibinfo {volume}
  {7}},\ \bibinfo {pages} {44024} (\bibinfo {year} {2017})}\BibitemShut
  {NoStop}%
\bibitem [{\citenamefont {Bourgeois-Hope}\ \emph {et~al.}(2016)\citenamefont
  {Bourgeois-Hope}, \citenamefont {Chi}, \citenamefont {Bonn}, \citenamefont
  {Liang}, \citenamefont {Hardy}, \citenamefont {Wolf}, \citenamefont
  {Meingast}, \citenamefont {Doiron-Leyraud},\ and\ \citenamefont
  {Taillefer}}]{hopePhysRevLett}%
  \BibitemOpen
  \bibfield  {author} {\bibinfo {author} {\bibfnamefont {P.}~\bibnamefont
  {Bourgeois-Hope}}, \bibinfo {author} {\bibfnamefont {S.}~\bibnamefont {Chi}},
  \bibinfo {author} {\bibfnamefont {D.~A.}\ \bibnamefont {Bonn}}, \bibinfo
  {author} {\bibfnamefont {R.}~\bibnamefont {Liang}}, \bibinfo {author}
  {\bibfnamefont {W.~N.}\ \bibnamefont {Hardy}}, \bibinfo {author}
  {\bibfnamefont {T.}~\bibnamefont {Wolf}}, \bibinfo {author} {\bibfnamefont
  {C.}~\bibnamefont {Meingast}}, \bibinfo {author} {\bibfnamefont
  {N.}~\bibnamefont {Doiron-Leyraud}}, \ and\ \bibinfo {author} {\bibfnamefont
  {L.}~\bibnamefont {Taillefer}},\ }\href@noop {} {\bibfield  {journal}
  {\bibinfo  {journal} {Phys. Rev. Lett.}\ }\textbf {\bibinfo {volume} {117}},\
  \bibinfo {pages} {097003} (\bibinfo {year} {2016})}\BibitemShut {NoStop}%
\bibitem [{\citenamefont {Abdel-Hafiez}\ \emph {et~al.}(2013)\citenamefont
  {Abdel-Hafiez}, \citenamefont {Ge}, \citenamefont {Vasiliev}, \citenamefont
  {Chareev}, \citenamefont {Van~de Vondel}, \citenamefont {Moshchalkov},\ and\
  \citenamefont {Silhanek}}]{AbdelHcFeSePRB}%
  \BibitemOpen
  \bibfield  {author} {\bibinfo {author} {\bibfnamefont {M.}~\bibnamefont
  {Abdel-Hafiez}}, \bibinfo {author} {\bibfnamefont {J.}~\bibnamefont {Ge}},
  \bibinfo {author} {\bibfnamefont {A.~N.}\ \bibnamefont {Vasiliev}}, \bibinfo
  {author} {\bibfnamefont {D.~A.}\ \bibnamefont {Chareev}}, \bibinfo {author}
  {\bibfnamefont {J.}~\bibnamefont {Van~de Vondel}}, \bibinfo {author}
  {\bibfnamefont {V.~V.}\ \bibnamefont {Moshchalkov}}, \ and\ \bibinfo {author}
  {\bibfnamefont {A.~V.}\ \bibnamefont {Silhanek}},\ }\href@noop {} {\bibfield
  {journal} {\bibinfo  {journal} {Phys. Rev. B}\ }\textbf {\bibinfo {volume}
  {88}},\ \bibinfo {pages} {174512} (\bibinfo {year} {2013})}\BibitemShut
  {NoStop}%
\bibitem [{\citenamefont {Sprau}\ \emph {et~al.}(2016)\citenamefont {Sprau},
  \citenamefont {Kostin}, \citenamefont {Kreisel}, \citenamefont {B{\"o}hmer},
  \citenamefont {Taufour}, \citenamefont {Canfield}, \citenamefont {Mukherjee},
  \citenamefont {Hirschfeld}, \citenamefont {Andersen},\ and\ \citenamefont
  {Davis}}]{BPQIarxiv}%
  \BibitemOpen
  \bibfield  {author} {\bibinfo {author} {\bibfnamefont {P.~O.}\ \bibnamefont
  {Sprau}}, \bibinfo {author} {\bibfnamefont {A.}~\bibnamefont {Kostin}},
  \bibinfo {author} {\bibfnamefont {A.}~\bibnamefont {Kreisel}}, \bibinfo
  {author} {\bibfnamefont {A.~E.}\ \bibnamefont {B{\"o}hmer}}, \bibinfo
  {author} {\bibfnamefont {V.}~\bibnamefont {Taufour}}, \bibinfo {author}
  {\bibfnamefont {P.~C.}\ \bibnamefont {Canfield}}, \bibinfo {author}
  {\bibfnamefont {S.}~\bibnamefont {Mukherjee}}, \bibinfo {author}
  {\bibfnamefont {P.~J.}\ \bibnamefont {Hirschfeld}}, \bibinfo {author}
  {\bibfnamefont {B.~M.}\ \bibnamefont {Andersen}}, \ and\ \bibinfo {author}
  {\bibfnamefont {J.~C.~S.}\ \bibnamefont {Davis}},\ }\href@noop {} {\bibfield
  {journal} {\bibinfo  {journal} {Science}\ }\textbf {\bibinfo {volume}
  {357}},\ \bibinfo {pages} {75} (\bibinfo {year} {2016})}\BibitemShut
  {NoStop}%
\bibitem [{\citenamefont {Dong}\ \emph {et~al.}(2009)\citenamefont {Dong},
  \citenamefont {Guan}, \citenamefont {Zhou}, \citenamefont {Qiu},
  \citenamefont {Ding}, \citenamefont {Zhang}, \citenamefont {Patel},
  \citenamefont {Xiao},\ and\ \citenamefont {Li}}]{FeSeoldthermalPRB}%
  \BibitemOpen
  \bibfield  {author} {\bibinfo {author} {\bibfnamefont {J.~K.}\ \bibnamefont
  {Dong}}, \bibinfo {author} {\bibfnamefont {T.~Y.}\ \bibnamefont {Guan}},
  \bibinfo {author} {\bibfnamefont {S.~Y.}\ \bibnamefont {Zhou}}, \bibinfo
  {author} {\bibfnamefont {X.}~\bibnamefont {Qiu}}, \bibinfo {author}
  {\bibfnamefont {L.}~\bibnamefont {Ding}}, \bibinfo {author} {\bibfnamefont
  {C.}~\bibnamefont {Zhang}}, \bibinfo {author} {\bibfnamefont
  {U.}~\bibnamefont {Patel}}, \bibinfo {author} {\bibfnamefont {Z.~L.}\
  \bibnamefont {Xiao}}, \ and\ \bibinfo {author} {\bibfnamefont {S.~Y.}\
  \bibnamefont {Li}},\ }\href@noop {} {\bibfield  {journal} {\bibinfo
  {journal} {Phys. Rev. B}\ }\textbf {\bibinfo {volume} {80}},\ \bibinfo
  {pages} {024518} (\bibinfo {year} {2009})}\BibitemShut {NoStop}%
\bibitem [{\citenamefont {Sun}\ \emph {et~al.}(2017{\natexlab{a}})\citenamefont
  {Sun}, \citenamefont {Kittaka}, \citenamefont {Nakamura}, \citenamefont
  {Sakakibara}, \citenamefont {Irie}, \citenamefont {Nomoto}, \citenamefont
  {Machida}, \citenamefont {Chen},\ and\ \citenamefont
  {Tamegai}}]{SunPRBARSHFeSe}%
  \BibitemOpen
  \bibfield  {author} {\bibinfo {author} {\bibfnamefont {Y.}~\bibnamefont
  {Sun}}, \bibinfo {author} {\bibfnamefont {S.}~\bibnamefont {Kittaka}},
  \bibinfo {author} {\bibfnamefont {S.}~\bibnamefont {Nakamura}}, \bibinfo
  {author} {\bibfnamefont {T.}~\bibnamefont {Sakakibara}}, \bibinfo {author}
  {\bibfnamefont {K.}~\bibnamefont {Irie}}, \bibinfo {author} {\bibfnamefont
  {T.}~\bibnamefont {Nomoto}}, \bibinfo {author} {\bibfnamefont
  {K.}~\bibnamefont {Machida}}, \bibinfo {author} {\bibfnamefont
  {J.}~\bibnamefont {Chen}}, \ and\ \bibinfo {author} {\bibfnamefont
  {T.}~\bibnamefont {Tamegai}},\ }\href@noop {} {\bibfield  {journal} {\bibinfo
   {journal} {Phys. Rev. B}\ }\textbf {\bibinfo {volume} {96}},\ \bibinfo
  {pages} {220505} (\bibinfo {year} {2017}{\natexlab{a}})}\BibitemShut
  {NoStop}%
\bibitem [{\citenamefont {Chen}\ \emph {et~al.}(2017)\citenamefont {Chen},
  \citenamefont {Zhu}, \citenamefont {Yang},\ and\ \citenamefont
  {Wen}}]{FeSeSHSecondJumpPhysRevB.96.064524}%
  \BibitemOpen
  \bibfield  {author} {\bibinfo {author} {\bibfnamefont {G.-Y.}\ \bibnamefont
  {Chen}}, \bibinfo {author} {\bibfnamefont {X.}~\bibnamefont {Zhu}}, \bibinfo
  {author} {\bibfnamefont {H.}~\bibnamefont {Yang}}, \ and\ \bibinfo {author}
  {\bibfnamefont {H.-H.}\ \bibnamefont {Wen}},\ }\href@noop {} {\bibfield
  {journal} {\bibinfo  {journal} {Phys. Rev. B}\ }\textbf {\bibinfo {volume}
  {96}},\ \bibinfo {pages} {064524} (\bibinfo {year} {2017})}\BibitemShut
  {NoStop}%
\bibitem [{\citenamefont {Kreisel}\ \emph {et~al.}(2015)\citenamefont
  {Kreisel}, \citenamefont {Mukherjee}, \citenamefont {Hirschfeld},\ and\
  \citenamefont {Andersen}}]{KreiselPRB}%
  \BibitemOpen
  \bibfield  {author} {\bibinfo {author} {\bibfnamefont {A.}~\bibnamefont
  {Kreisel}}, \bibinfo {author} {\bibfnamefont {S.}~\bibnamefont {Mukherjee}},
  \bibinfo {author} {\bibfnamefont {P.~J.}\ \bibnamefont {Hirschfeld}}, \ and\
  \bibinfo {author} {\bibfnamefont {B.~M.}\ \bibnamefont {Andersen}},\
  }\href@noop {} {\bibfield  {journal} {\bibinfo  {journal} {Phys. Rev. B}\
  }\textbf {\bibinfo {volume} {92}},\ \bibinfo {pages} {224515} (\bibinfo
  {year} {2015})}\BibitemShut {NoStop}%
\bibitem [{\citenamefont {B\"ohmer}\ \emph {et~al.}(2016)\citenamefont
  {B\"ohmer}, \citenamefont {Taufour}, \citenamefont {Straszheim},
  \citenamefont {Wolf},\ and\ \citenamefont
  {Canfield}}]{BohmerdisorderPhysRevB.94.024526}%
  \BibitemOpen
  \bibfield  {author} {\bibinfo {author} {\bibfnamefont {A.~E.}\ \bibnamefont
  {B\"ohmer}}, \bibinfo {author} {\bibfnamefont {V.}~\bibnamefont {Taufour}},
  \bibinfo {author} {\bibfnamefont {W.~E.}\ \bibnamefont {Straszheim}},
  \bibinfo {author} {\bibfnamefont {T.}~\bibnamefont {Wolf}}, \ and\ \bibinfo
  {author} {\bibfnamefont {P.~C.}\ \bibnamefont {Canfield}},\ }\href {\doibase
  10.1103/PhysRevB.94.024526} {\bibfield  {journal} {\bibinfo  {journal} {Phys.
  Rev. B}\ }\textbf {\bibinfo {volume} {94}},\ \bibinfo {pages} {024526}
  (\bibinfo {year} {2016})}\BibitemShut {NoStop}%
\bibitem [{\citenamefont {R\"o\ss{}ler}\ \emph {et~al.}(2018)\citenamefont
  {R\"o\ss{}ler}, \citenamefont {Huang}, \citenamefont {Jiao}, \citenamefont
  {Koz}, \citenamefont {Schwarz},\ and\ \citenamefont
  {Wirth}}]{RosslerPhysRevB.97.094503}%
  \BibitemOpen
  \bibfield  {author} {\bibinfo {author} {\bibfnamefont {S.}~\bibnamefont
  {R\"o\ss{}ler}}, \bibinfo {author} {\bibfnamefont {C.-L.}\ \bibnamefont
  {Huang}}, \bibinfo {author} {\bibfnamefont {L.}~\bibnamefont {Jiao}},
  \bibinfo {author} {\bibfnamefont {C.}~\bibnamefont {Koz}}, \bibinfo {author}
  {\bibfnamefont {U.}~\bibnamefont {Schwarz}}, \ and\ \bibinfo {author}
  {\bibfnamefont {S.}~\bibnamefont {Wirth}},\ }\href {\doibase
  10.1103/PhysRevB.97.094503} {\bibfield  {journal} {\bibinfo  {journal} {Phys.
  Rev. B}\ }\textbf {\bibinfo {volume} {97}},\ \bibinfo {pages} {094503}
  (\bibinfo {year} {2018})}\BibitemShut {NoStop}%
\bibitem [{\citenamefont {Teknowijoyo}\ \emph {et~al.}(2016)\citenamefont
  {Teknowijoyo}, \citenamefont {Cho}, \citenamefont {Tanatar}, \citenamefont
  {Gonzales}, \citenamefont {B\"ohmer}, \citenamefont {Cavani}, \citenamefont
  {Mishra}, \citenamefont {Hirschfeld}, \citenamefont {Bud'ko}, \citenamefont
  {Canfield},\ and\ \citenamefont {Prozorov}}]{FeSeelectronIrra}%
  \BibitemOpen
  \bibfield  {author} {\bibinfo {author} {\bibfnamefont {S.}~\bibnamefont
  {Teknowijoyo}}, \bibinfo {author} {\bibfnamefont {K.}~\bibnamefont {Cho}},
  \bibinfo {author} {\bibfnamefont {M.~A.}\ \bibnamefont {Tanatar}}, \bibinfo
  {author} {\bibfnamefont {J.}~\bibnamefont {Gonzales}}, \bibinfo {author}
  {\bibfnamefont {A.~E.}\ \bibnamefont {B\"ohmer}}, \bibinfo {author}
  {\bibfnamefont {O.}~\bibnamefont {Cavani}}, \bibinfo {author} {\bibfnamefont
  {V.}~\bibnamefont {Mishra}}, \bibinfo {author} {\bibfnamefont {P.~J.}\
  \bibnamefont {Hirschfeld}}, \bibinfo {author} {\bibfnamefont {S.~L.}\
  \bibnamefont {Bud'ko}}, \bibinfo {author} {\bibfnamefont {P.~C.}\
  \bibnamefont {Canfield}}, \ and\ \bibinfo {author} {\bibfnamefont
  {R.}~\bibnamefont {Prozorov}},\ }\href@noop {} {\bibfield  {journal}
  {\bibinfo  {journal} {Phys. Rev. B}\ }\textbf {\bibinfo {volume} {94}},\
  \bibinfo {pages} {064521} (\bibinfo {year} {2016})}\BibitemShut {NoStop}%
\bibitem [{\citenamefont {Sun}\ \emph {et~al.}(2017{\natexlab{b}})\citenamefont
  {Sun}, \citenamefont {Park}, \citenamefont {Pyon}, \citenamefont {Tamegai},\
  and\ \citenamefont {Kitamura}}]{SunPhysRevBprotonirra}%
  \BibitemOpen
  \bibfield  {author} {\bibinfo {author} {\bibfnamefont {Y.}~\bibnamefont
  {Sun}}, \bibinfo {author} {\bibfnamefont {A.}~\bibnamefont {Park}}, \bibinfo
  {author} {\bibfnamefont {S.}~\bibnamefont {Pyon}}, \bibinfo {author}
  {\bibfnamefont {T.}~\bibnamefont {Tamegai}}, \ and\ \bibinfo {author}
  {\bibfnamefont {H.}~\bibnamefont {Kitamura}},\ }\href@noop {} {\bibfield
  {journal} {\bibinfo  {journal} {Phys. Rev. B}\ }\textbf {\bibinfo {volume}
  {96}},\ \bibinfo {pages} {140505} (\bibinfo {year}
  {2017}{\natexlab{b}})}\BibitemShut {NoStop}%
\bibitem [{\citenamefont {Sun}\ \emph {et~al.}(2017{\natexlab{c}})\citenamefont
  {Sun}, \citenamefont {Park}, \citenamefont {Pyon}, \citenamefont {Tamegai},
  \citenamefont {Kambara},\ and\ \citenamefont
  {Ichinose}}]{SunPRBheavyionIrradiation}%
  \BibitemOpen
  \bibfield  {author} {\bibinfo {author} {\bibfnamefont {Y.}~\bibnamefont
  {Sun}}, \bibinfo {author} {\bibfnamefont {A.}~\bibnamefont {Park}}, \bibinfo
  {author} {\bibfnamefont {S.}~\bibnamefont {Pyon}}, \bibinfo {author}
  {\bibfnamefont {T.}~\bibnamefont {Tamegai}}, \bibinfo {author} {\bibfnamefont
  {T.}~\bibnamefont {Kambara}}, \ and\ \bibinfo {author} {\bibfnamefont
  {A.}~\bibnamefont {Ichinose}},\ }\href@noop {} {\bibfield  {journal}
  {\bibinfo  {journal} {Phys. Rev. B}\ }\textbf {\bibinfo {volume} {95}},\
  \bibinfo {pages} {104514} (\bibinfo {year} {2017}{\natexlab{c}})}\BibitemShut
  {NoStop}%
\bibitem [{\citenamefont {Sakakibara}\ \emph {et~al.}(2016)\citenamefont
  {Sakakibara}, \citenamefont {Kittaka},\ and\ \citenamefont
  {Machida}}]{SakakibaraReview}%
  \BibitemOpen
  \bibfield  {author} {\bibinfo {author} {\bibfnamefont {T.}~\bibnamefont
  {Sakakibara}}, \bibinfo {author} {\bibfnamefont {S.}~\bibnamefont {Kittaka}},
  \ and\ \bibinfo {author} {\bibfnamefont {K.}~\bibnamefont {Machida}},\
  }\href@noop {} {\bibfield  {journal} {\bibinfo  {journal} {Rep. Prog. Phys.}\
  }\textbf {\bibinfo {volume} {79}},\ \bibinfo {pages} {094002} (\bibinfo
  {year} {2016})}\BibitemShut {NoStop}%
\bibitem [{\citenamefont {Sun}\ \emph {et~al.}(2015)\citenamefont {Sun},
  \citenamefont {Pyon}, \citenamefont {Tamegai}, \citenamefont {Kobayashi},
  \citenamefont {Watashige}, \citenamefont {Kasahara}, \citenamefont
  {Matsuda},\ and\ \citenamefont {Shibauchi}}]{SunPhysRevBJcFeSe}%
  \BibitemOpen
  \bibfield  {author} {\bibinfo {author} {\bibfnamefont {Y.}~\bibnamefont
  {Sun}}, \bibinfo {author} {\bibfnamefont {S.}~\bibnamefont {Pyon}}, \bibinfo
  {author} {\bibfnamefont {T.}~\bibnamefont {Tamegai}}, \bibinfo {author}
  {\bibfnamefont {R.}~\bibnamefont {Kobayashi}}, \bibinfo {author}
  {\bibfnamefont {T.}~\bibnamefont {Watashige}}, \bibinfo {author}
  {\bibfnamefont {S.}~\bibnamefont {Kasahara}}, \bibinfo {author}
  {\bibfnamefont {Y.}~\bibnamefont {Matsuda}}, \ and\ \bibinfo {author}
  {\bibfnamefont {T.}~\bibnamefont {Shibauchi}},\ }\href {\doibase
  10.1103/PhysRevB.92.144509} {\bibfield  {journal} {\bibinfo  {journal} {Phys.
  Rev. B}\ }\textbf {\bibinfo {volume} {92}},\ \bibinfo {pages} {144509}
  (\bibinfo {year} {2015})}\BibitemShut {NoStop}%
\bibitem [{\citenamefont {Nakayama}\ \emph {et~al.}(2014)\citenamefont
  {Nakayama}, \citenamefont {Miyata}, \citenamefont {Phan}, \citenamefont
  {Sato}, \citenamefont {Tanabe}, \citenamefont {Urata}, \citenamefont
  {Tanigaki},\ and\ \citenamefont {Takahashi}}]{NakayamaPRL}%
  \BibitemOpen
  \bibfield  {author} {\bibinfo {author} {\bibfnamefont {K.}~\bibnamefont
  {Nakayama}}, \bibinfo {author} {\bibfnamefont {Y.}~\bibnamefont {Miyata}},
  \bibinfo {author} {\bibfnamefont {G.~N.}\ \bibnamefont {Phan}}, \bibinfo
  {author} {\bibfnamefont {T.}~\bibnamefont {Sato}}, \bibinfo {author}
  {\bibfnamefont {Y.}~\bibnamefont {Tanabe}}, \bibinfo {author} {\bibfnamefont
  {T.}~\bibnamefont {Urata}}, \bibinfo {author} {\bibfnamefont
  {K.}~\bibnamefont {Tanigaki}}, \ and\ \bibinfo {author} {\bibfnamefont
  {T.}~\bibnamefont {Takahashi}},\ }\href@noop {} {\bibfield  {journal}
  {\bibinfo  {journal} {Phys. Rev. Lett.}\ }\textbf {\bibinfo {volume} {113}},\
  \bibinfo {pages} {237001} (\bibinfo {year} {2014})}\BibitemShut {NoStop}%
\bibitem [{\citenamefont {Watson}\ \emph {et~al.}(2015)\citenamefont {Watson},
  \citenamefont {Kim}, \citenamefont {Haghighirad}, \citenamefont {Blake},
  \citenamefont {Davies}, \citenamefont {Hoesch}, \citenamefont {Wolf},\ and\
  \citenamefont {Coldea}}]{WatsonPRB91}%
  \BibitemOpen
  \bibfield  {author} {\bibinfo {author} {\bibfnamefont {M.~D.}\ \bibnamefont
  {Watson}}, \bibinfo {author} {\bibfnamefont {T.~K.}\ \bibnamefont {Kim}},
  \bibinfo {author} {\bibfnamefont {A.~A.}\ \bibnamefont {Haghighirad}},
  \bibinfo {author} {\bibfnamefont {S.~F.}\ \bibnamefont {Blake}}, \bibinfo
  {author} {\bibfnamefont {N.~R.}\ \bibnamefont {Davies}}, \bibinfo {author}
  {\bibfnamefont {M.}~\bibnamefont {Hoesch}}, \bibinfo {author} {\bibfnamefont
  {T.}~\bibnamefont {Wolf}}, \ and\ \bibinfo {author} {\bibfnamefont {A.~I.}\
  \bibnamefont {Coldea}},\ }\href@noop {} {\bibfield  {journal} {\bibinfo
  {journal} {Phys. Rev. B}\ }\textbf {\bibinfo {volume} {92}},\ \bibinfo
  {pages} {121108} (\bibinfo {year} {2015})}\BibitemShut {NoStop}%
\bibitem [{\citenamefont {Watson}\ \emph {et~al.}(2016)\citenamefont {Watson},
  \citenamefont {Kim}, \citenamefont {Rhodes}, \citenamefont {Eschrig},
  \citenamefont {Hoesch}, \citenamefont {Haghighirad},\ and\ \citenamefont
  {Coldea}}]{WatsonPRB2016}%
  \BibitemOpen
  \bibfield  {author} {\bibinfo {author} {\bibfnamefont {M.~D.}\ \bibnamefont
  {Watson}}, \bibinfo {author} {\bibfnamefont {T.~K.}\ \bibnamefont {Kim}},
  \bibinfo {author} {\bibfnamefont {L.~C.}\ \bibnamefont {Rhodes}}, \bibinfo
  {author} {\bibfnamefont {M.}~\bibnamefont {Eschrig}}, \bibinfo {author}
  {\bibfnamefont {M.}~\bibnamefont {Hoesch}}, \bibinfo {author} {\bibfnamefont
  {A.~A.}\ \bibnamefont {Haghighirad}}, \ and\ \bibinfo {author} {\bibfnamefont
  {A.~I.}\ \bibnamefont {Coldea}},\ }\href@noop {} {\bibfield  {journal}
  {\bibinfo  {journal} {Phys. Rev. B}\ }\textbf {\bibinfo {volume} {94}},\
  \bibinfo {pages} {201107} (\bibinfo {year} {2016})}\BibitemShut {NoStop}%
\bibitem [{\citenamefont {Coldea}\ and\ \citenamefont
  {Watson}(2018)}]{Coldeaannurev}%
  \BibitemOpen
  \bibfield  {author} {\bibinfo {author} {\bibfnamefont {A.~I.}\ \bibnamefont
  {Coldea}}\ and\ \bibinfo {author} {\bibfnamefont {M.~D.}\ \bibnamefont
  {Watson}},\ }\href@noop {} {\bibfield  {journal} {\bibinfo  {journal} {Annu.
  Rev. Condens. Matter Phys.}\ }\textbf {\bibinfo {volume} {9}},\ \bibinfo
  {pages} {125} (\bibinfo {year} {2018})}\BibitemShut {NoStop}%
\bibitem [{\citenamefont {Fedorov}\ \emph {et~al.}(2016)\citenamefont
  {Fedorov}, \citenamefont {Yaresko}, \citenamefont {Kim}, \citenamefont
  {Kushnirenko}, \citenamefont {Haubold}, \citenamefont {Wolf}, \citenamefont
  {Hoesch}, \citenamefont {Grüneis}, \citenamefont {Büchner},\ and\
  \citenamefont {Borisenko}}]{FedorovARPESSciRep}%
  \BibitemOpen
  \bibfield  {author} {\bibinfo {author} {\bibfnamefont {A.}~\bibnamefont
  {Fedorov}}, \bibinfo {author} {\bibfnamefont {A.}~\bibnamefont {Yaresko}},
  \bibinfo {author} {\bibfnamefont {T.~K.}\ \bibnamefont {Kim}}, \bibinfo
  {author} {\bibfnamefont {Y.}~\bibnamefont {Kushnirenko}}, \bibinfo {author}
  {\bibfnamefont {E.}~\bibnamefont {Haubold}}, \bibinfo {author} {\bibfnamefont
  {T.}~\bibnamefont {Wolf}}, \bibinfo {author} {\bibfnamefont {M.}~\bibnamefont
  {Hoesch}}, \bibinfo {author} {\bibfnamefont {A.}~\bibnamefont {Grüneis}},
  \bibinfo {author} {\bibfnamefont {B.}~\bibnamefont {Büchner}}, \ and\
  \bibinfo {author} {\bibfnamefont {S.~V.}\ \bibnamefont {Borisenko}},\
  }\href@noop {} {\bibfield  {journal} {\bibinfo  {journal} {Sci. Rep.}\
  }\textbf {\bibinfo {volume} {6}},\ \bibinfo {pages} {36834} (\bibinfo {year}
  {2016})}\BibitemShut {NoStop}%
\bibitem [{\citenamefont {Jiao}\ \emph {et~al.}(2017)\citenamefont {Jiao},
  \citenamefont {R\"o\ss{}ler}, \citenamefont {Koz}, \citenamefont {Schwarz},
  \citenamefont {Kasinathan}, \citenamefont {R\"o\ss{}ler},\ and\ \citenamefont
  {Wirth}}]{LinJiaoSTMFevacancyFeSePhysRevB.96.094504}%
  \BibitemOpen
  \bibfield  {author} {\bibinfo {author} {\bibfnamefont {L.}~\bibnamefont
  {Jiao}}, \bibinfo {author} {\bibfnamefont {S.}~\bibnamefont {R\"o\ss{}ler}},
  \bibinfo {author} {\bibfnamefont {C.}~\bibnamefont {Koz}}, \bibinfo {author}
  {\bibfnamefont {U.}~\bibnamefont {Schwarz}}, \bibinfo {author} {\bibfnamefont
  {D.}~\bibnamefont {Kasinathan}}, \bibinfo {author} {\bibfnamefont {U.~K.}\
  \bibnamefont {R\"o\ss{}ler}}, \ and\ \bibinfo {author} {\bibfnamefont
  {S.}~\bibnamefont {Wirth}},\ }\href@noop {} {\bibfield  {journal} {\bibinfo
  {journal} {Phys. Rev. B}\ }\textbf {\bibinfo {volume} {96}},\ \bibinfo
  {pages} {094504} (\bibinfo {year} {2017})}\BibitemShut {NoStop}%
\bibitem [{\citenamefont {Hashimoto}\ \emph {et~al.}(2018)\citenamefont
  {Hashimoto}, \citenamefont {Ota}, \citenamefont {Yamamoto}, \citenamefont
  {Suzuki}, \citenamefont {Shimojima}, \citenamefont {Watanabe}, \citenamefont
  {Chen}, \citenamefont {Kasahara}, \citenamefont {Matsuda}, \citenamefont
  {Shibauchi}, \citenamefont {Okazaki},\ and\ \citenamefont
  {Shin}}]{Hashimotonatcomm}%
  \BibitemOpen
  \bibfield  {author} {\bibinfo {author} {\bibfnamefont {T.}~\bibnamefont
  {Hashimoto}}, \bibinfo {author} {\bibfnamefont {Y.}~\bibnamefont {Ota}},
  \bibinfo {author} {\bibfnamefont {H.~Q.}\ \bibnamefont {Yamamoto}}, \bibinfo
  {author} {\bibfnamefont {Y.}~\bibnamefont {Suzuki}}, \bibinfo {author}
  {\bibfnamefont {T.}~\bibnamefont {Shimojima}}, \bibinfo {author}
  {\bibfnamefont {S.}~\bibnamefont {Watanabe}}, \bibinfo {author}
  {\bibfnamefont {C.}~\bibnamefont {Chen}}, \bibinfo {author} {\bibfnamefont
  {S.}~\bibnamefont {Kasahara}}, \bibinfo {author} {\bibfnamefont
  {Y.}~\bibnamefont {Matsuda}}, \bibinfo {author} {\bibfnamefont
  {T.}~\bibnamefont {Shibauchi}}, \bibinfo {author} {\bibfnamefont
  {K.}~\bibnamefont {Okazaki}}, \ and\ \bibinfo {author} {\bibfnamefont
  {S.}~\bibnamefont {Shin}},\ }\href@noop {} {\bibfield  {journal} {\bibinfo
  {journal} {Nat. Commun.}\ }\textbf {\bibinfo {volume} {9}},\ \bibinfo {pages}
  {282} (\bibinfo {year} {2018})}\BibitemShut {NoStop}%
\bibitem [{\citenamefont {Bouquet}\ \emph {et~al.}(2002)\citenamefont
  {Bouquet}, \citenamefont {Wang}, \citenamefont {Sheikin}, \citenamefont
  {Plackowski}, \citenamefont {Junod}, \citenamefont {Lee},\ and\ \citenamefont
  {Tajima}}]{MgB2PhysRevLetttwogap}%
  \BibitemOpen
  \bibfield  {author} {\bibinfo {author} {\bibfnamefont {F.}~\bibnamefont
  {Bouquet}}, \bibinfo {author} {\bibfnamefont {Y.}~\bibnamefont {Wang}},
  \bibinfo {author} {\bibfnamefont {I.}~\bibnamefont {Sheikin}}, \bibinfo
  {author} {\bibfnamefont {T.}~\bibnamefont {Plackowski}}, \bibinfo {author}
  {\bibfnamefont {A.}~\bibnamefont {Junod}}, \bibinfo {author} {\bibfnamefont
  {S.}~\bibnamefont {Lee}}, \ and\ \bibinfo {author} {\bibfnamefont
  {S.}~\bibnamefont {Tajima}},\ }\href@noop {} {\bibfield  {journal} {\bibinfo
  {journal} {Phys. Rev. Lett.}\ }\textbf {\bibinfo {volume} {89}},\ \bibinfo
  {pages} {257001} (\bibinfo {year} {2002})}\BibitemShut {NoStop}%
\bibitem [{\citenamefont {Nakajima}\ \emph {et~al.}(2008)\citenamefont
  {Nakajima}, \citenamefont {Nakagawa}, \citenamefont {Tamegai},\ and\
  \citenamefont {Harima}}]{NakajimaRuFeSitwogapPhysRevLett.100.157001}%
  \BibitemOpen
  \bibfield  {author} {\bibinfo {author} {\bibfnamefont {Y.}~\bibnamefont
  {Nakajima}}, \bibinfo {author} {\bibfnamefont {T.}~\bibnamefont {Nakagawa}},
  \bibinfo {author} {\bibfnamefont {T.}~\bibnamefont {Tamegai}}, \ and\
  \bibinfo {author} {\bibfnamefont {H.}~\bibnamefont {Harima}},\ }\href@noop {}
  {\bibfield  {journal} {\bibinfo  {journal} {Phys. Rev. Lett.}\ }\textbf
  {\bibinfo {volume} {100}},\ \bibinfo {pages} {157001} (\bibinfo {year}
  {2008})}\BibitemShut {NoStop}%
\bibitem [{\citenamefont {Hardy}\ \emph {et~al.}(2018)\citenamefont {Hardy},
  \citenamefont {He}, \citenamefont {Wang}, \citenamefont {Wolf}, \citenamefont
  {Schweiss}, \citenamefont {Merz}, \citenamefont {Barth}, \citenamefont
  {Adelmann}, \citenamefont {Eder}, \citenamefont {Haghighirad},\ and\
  \citenamefont {Meingast}}]{Hardyarxiv}%
  \BibitemOpen
  \bibfield  {author} {\bibinfo {author} {\bibfnamefont {F.}~\bibnamefont
  {Hardy}}, \bibinfo {author} {\bibfnamefont {M.}~\bibnamefont {He}}, \bibinfo
  {author} {\bibfnamefont {L.}~\bibnamefont {Wang}}, \bibinfo {author}
  {\bibfnamefont {T.}~\bibnamefont {Wolf}}, \bibinfo {author} {\bibfnamefont
  {P.}~\bibnamefont {Schweiss}}, \bibinfo {author} {\bibfnamefont
  {M.}~\bibnamefont {Merz}}, \bibinfo {author} {\bibfnamefont {M.}~\bibnamefont
  {Barth}}, \bibinfo {author} {\bibfnamefont {P.}~\bibnamefont {Adelmann}},
  \bibinfo {author} {\bibfnamefont {R.}~\bibnamefont {Eder}}, \bibinfo {author}
  {\bibfnamefont {A.-A.}\ \bibnamefont {Haghighirad}}, \ and\ \bibinfo {author}
  {\bibfnamefont {C.}~\bibnamefont {Meingast}},\ }\href@noop {} {\bibfield
  {journal} {\bibinfo  {journal} {arXiv:1807.07907}\ } (\bibinfo {year}
  {2018})}\BibitemShut {NoStop}%
\bibitem [{\citenamefont {Sato}\ \emph {et~al.}(2017)\citenamefont {Sato},
  \citenamefont {Kasahara}, \citenamefont {Taniguchi}, \citenamefont {Xing},
  \citenamefont {Kasahara}, \citenamefont {Tokiwa}, \citenamefont {Yamakawa},
  \citenamefont {Kontani}, \citenamefont {Shibauchi},\ and\ \citenamefont
  {Matsuda}}]{SatoFeSeSPNAS}%
  \BibitemOpen
  \bibfield  {author} {\bibinfo {author} {\bibfnamefont {Y.}~\bibnamefont
  {Sato}}, \bibinfo {author} {\bibfnamefont {S.}~\bibnamefont {Kasahara}},
  \bibinfo {author} {\bibfnamefont {T.}~\bibnamefont {Taniguchi}}, \bibinfo
  {author} {\bibfnamefont {X.}~\bibnamefont {Xing}}, \bibinfo {author}
  {\bibfnamefont {Y.}~\bibnamefont {Kasahara}}, \bibinfo {author}
  {\bibfnamefont {Y.}~\bibnamefont {Tokiwa}}, \bibinfo {author} {\bibfnamefont
  {Y.}~\bibnamefont {Yamakawa}}, \bibinfo {author} {\bibfnamefont
  {H.}~\bibnamefont {Kontani}}, \bibinfo {author} {\bibfnamefont
  {T.}~\bibnamefont {Shibauchi}}, \ and\ \bibinfo {author} {\bibfnamefont
  {Y.}~\bibnamefont {Matsuda}},\ }\href@noop {} {\bibfield  {journal} {\bibinfo
   {journal} {arXiv:1705.09074}\ } (\bibinfo {year} {2017})}\BibitemShut
  {NoStop}%
\bibitem [{\citenamefont {Volovik}(1993)}]{VolovikJETPLett}%
  \BibitemOpen
  \bibfield  {author} {\bibinfo {author} {\bibfnamefont {G.~E.}\ \bibnamefont
  {Volovik}},\ }\href@noop {} {\bibfield  {journal} {\bibinfo  {journal} {JETP
  Lett.}\ }\textbf {\bibinfo {volume} {58}},\ \bibinfo {pages} {469} (\bibinfo
  {year} {1993})}\BibitemShut {NoStop}%
\bibitem [{\citenamefont {Vorontsov}\ and\ \citenamefont
  {Vekhter}(2006)}]{VorontsovPhysRevLett.96.237001}%
  \BibitemOpen
  \bibfield  {author} {\bibinfo {author} {\bibfnamefont {A.}~\bibnamefont
  {Vorontsov}}\ and\ \bibinfo {author} {\bibfnamefont {I.}~\bibnamefont
  {Vekhter}},\ }\href {\doibase 10.1103/PhysRevLett.96.237001} {\bibfield
  {journal} {\bibinfo  {journal} {Phys. Rev. Lett.}\ }\textbf {\bibinfo
  {volume} {96}},\ \bibinfo {pages} {237001} (\bibinfo {year}
  {2006})}\BibitemShut {NoStop}%
\bibitem [{\citenamefont {Hiragi}\ \emph {et~al.}(2010)\citenamefont {Hiragi},
  \citenamefont {Suzuki}, \citenamefont {Ichioka},\ and\ \citenamefont
  {Machida}}]{HiragiJPSJASHcal}%
  \BibitemOpen
  \bibfield  {author} {\bibinfo {author} {\bibfnamefont {M.}~\bibnamefont
  {Hiragi}}, \bibinfo {author} {\bibfnamefont {K.~M.}\ \bibnamefont {Suzuki}},
  \bibinfo {author} {\bibfnamefont {M.}~\bibnamefont {Ichioka}}, \ and\
  \bibinfo {author} {\bibfnamefont {K.}~\bibnamefont {Machida}},\ }\href@noop
  {} {\bibfield  {journal} {\bibinfo  {journal} {J. Phys. Soc. Jpn.}\ }\textbf
  {\bibinfo {volume} {79}},\ \bibinfo {pages} {094709} (\bibinfo {year}
  {2010})}\BibitemShut {NoStop}%
\bibitem [{\citenamefont {An}\ \emph {et~al.}(2010)\citenamefont {An},
  \citenamefont {Sakakibara}, \citenamefont {Settai}, \citenamefont {Onuki},
  \citenamefont {Hiragi}, \citenamefont {Ichioka},\ and\ \citenamefont
  {Machida}}]{AnPhysRevLett.104.037002}%
  \BibitemOpen
  \bibfield  {author} {\bibinfo {author} {\bibfnamefont {K.}~\bibnamefont
  {An}}, \bibinfo {author} {\bibfnamefont {T.}~\bibnamefont {Sakakibara}},
  \bibinfo {author} {\bibfnamefont {R.}~\bibnamefont {Settai}}, \bibinfo
  {author} {\bibfnamefont {Y.}~\bibnamefont {Onuki}}, \bibinfo {author}
  {\bibfnamefont {M.}~\bibnamefont {Hiragi}}, \bibinfo {author} {\bibfnamefont
  {M.}~\bibnamefont {Ichioka}}, \ and\ \bibinfo {author} {\bibfnamefont
  {K.}~\bibnamefont {Machida}},\ }\href@noop {} {\bibfield  {journal} {\bibinfo
   {journal} {Phys. Rev. Lett.}\ }\textbf {\bibinfo {volume} {104}},\ \bibinfo
  {pages} {037002} (\bibinfo {year} {2010})}\BibitemShut {NoStop}%
\bibitem [{\citenamefont {Kittaka}\ \emph {et~al.}(2016)\citenamefont
  {Kittaka}, \citenamefont {Shimizu}, \citenamefont {Sakakibara}, \citenamefont
  {Haga}, \citenamefont {Yamamoto}, \citenamefont {\={O}nuki}, \citenamefont
  {Tsutsumi}, \citenamefont {Nomoto}, \citenamefont {Ikeda},\ and\
  \citenamefont {Machida}}]{KittakaKFe2As2JPSJ}%
  \BibitemOpen
  \bibfield  {author} {\bibinfo {author} {\bibfnamefont {S.}~\bibnamefont
  {Kittaka}}, \bibinfo {author} {\bibfnamefont {Y.}~\bibnamefont {Shimizu}},
  \bibinfo {author} {\bibfnamefont {T.}~\bibnamefont {Sakakibara}}, \bibinfo
  {author} {\bibfnamefont {Y.}~\bibnamefont {Haga}}, \bibinfo {author}
  {\bibfnamefont {E.}~\bibnamefont {Yamamoto}}, \bibinfo {author}
  {\bibfnamefont {Y.}~\bibnamefont {\={O}nuki}}, \bibinfo {author}
  {\bibfnamefont {Y.}~\bibnamefont {Tsutsumi}}, \bibinfo {author}
  {\bibfnamefont {T.}~\bibnamefont {Nomoto}}, \bibinfo {author} {\bibfnamefont
  {H.}~\bibnamefont {Ikeda}}, \ and\ \bibinfo {author} {\bibfnamefont
  {K.}~\bibnamefont {Machida}},\ }\href {\doibase 10.7566/JPSJ.85.033704}
  {\bibfield  {journal} {\bibinfo  {journal} {J. Phys. Soc. Jpn.}\ }\textbf
  {\bibinfo {volume} {85}},\ \bibinfo {pages} {033704} (\bibinfo {year}
  {2016})}\BibitemShut {NoStop}%
\bibitem [{\citenamefont {Kittaka}\ \emph {et~al.}(2013)\citenamefont
  {Kittaka}, \citenamefont {Sakakibara}, \citenamefont {Hedo}, \citenamefont
  {\={O}nuki},\ and\ \citenamefont {Machida}}]{KittakaCeRu2JPSJ.82.123706}%
  \BibitemOpen
  \bibfield  {author} {\bibinfo {author} {\bibfnamefont {S.}~\bibnamefont
  {Kittaka}}, \bibinfo {author} {\bibfnamefont {T.}~\bibnamefont {Sakakibara}},
  \bibinfo {author} {\bibfnamefont {M.}~\bibnamefont {Hedo}}, \bibinfo {author}
  {\bibfnamefont {Y.}~\bibnamefont {\={O}nuki}}, \ and\ \bibinfo {author}
  {\bibfnamefont {K.}~\bibnamefont {Machida}},\ }\href {\doibase
  10.7566/JPSJ.82.123706} {\bibfield  {journal} {\bibinfo  {journal} {J. Phys.
  Soc. Jpn.}\ }\textbf {\bibinfo {volume} {82}},\ \bibinfo {pages} {123706}
  (\bibinfo {year} {2013})}\BibitemShut {NoStop}%
\end{thebibliography}%

\end{document}